\documentstyle[aps,floats,epsfig,fancyheadings]{revtex}
\def\YBCO {$\rm{YBa_2Cu_3O_{7-\delta}}$}
\def\SmBCO {$\rm{SmBa_2Cu_3O_{7-\delta}}$}
\begin{document}
\headrulewidth=0pt
\chead{\small The following article has been submitted to Applied Physics Letters.\\ After it is published, it will be found at http://ojps.aip.org/aplo/.}
\cfoot{\thepage}
\draft

% uncomment the following two lines for two-column version
\twocolumn[\hsize\textwidth\columnwidth\hsize\csname
@twocolumnfalse\endcsname

\preprint{Gruss et al. APL}
\title{Superconducting Bulk Magnets: Very High Trapped Fields and Cracking}
\author{S. Gruss, G. Fuchs, G. Krabbes, P. Verges, G. St\"{o}ver, K.-H. M\"{u}ller, J. Fink and L. Schultz}
\address{Institut f\"{u}r Festk\"{o}rper- und Werkstofforschung Dresden,
P.O. Box 27 01 16, D-01171 Dresden, Germany}
\date{04/20/2001}
\maketitle
\thispagestyle{fancy}

\begin{abstract}
Improved trapped fields are reported for bulk melt-textured \YBCO \ (YBCO) material in the temperature range between 20\,K and 50\,K. Trapped fields up to 12.2\,T were obtained at 22\,K on the surface of single YBCO disks (with Ag and Zn additions). In YBCO mini-magnets, maximum trapped fields of 16\,T (at 24\,K) and of 11.2\,T (at 47\,K) were achieved using (Zn + Ag) and Zn additions, respectively. In all cases, the YBCO disks were encapsulated in steel tubes in order to reinforce the material against the large tensile stress acting during the magnetizing process and to avoid cracking. We observed cracking not only during the magnetizing process, but also as a consequence of flux jumps due to thermomagnetic instabilities in the temperature range betweeen 20\,K and 30\,K.
\end{abstract}
\pacs{7460Ge, 7462Bf, 7480Bj}

% uncomment the following line for two-column version
\vskip2pc]

\narrowtext
Bulk type II superconductors can trap high magnetic fields that are generated by superconducting persistent currents circulating macroscopically within the superconductor. The main features of the resulting field distribution are a maximum trapped field $B_{\rm 0}$ in the center of the superconducting domain and a field gradient towards the sample edge which is determined by the critical current density $j_{\rm c}$ of the supercurrents. Therefore, high trapped fields $B_{\rm 0}$ in bulk superconductors require a high critical current density and a large size of the current loops. Large single-domain bulk \YBCO \ based (YBCO) samples can be produced by melt texture processing, especially by using \SmBCO \ (Sm-123) as a seed crystal.\cite{1} The pinning effect in melt textured samples can be improved by irradiation methods\cite{2,3} and alternatively, by zinc-doping.\cite{4,5} Cracking of the samples was found to limit the trapped field of bulk YBCO at temperatures below 77\,K which can be explained by tensile stresses that occur during the magnetization process due to the stored flux density gradient and may exceed the tensile strength of the material.\cite{6} The mechanical properties of bulk YBCO and its tensile strength can be improved considerably by the addition of Ag.\cite{7,8} Another possibility to avoid cracking during magnetizing is to encapsulate the bulk YBCO disks in steel tubes.\cite{9} A steel tube leads to stress compensation by generating a compressive stress on YBCO after cooling from 300\,K to the measuring temperature which is due to the higher thermal expansion coefficient of steel compared to that of YBCO in the $a,b$-plane.
Maximum trapped fields $B_{\rm 0}$ of 11.5\,T at 17\,K and of 14.4\,T at 22\,K were reported for single YBCO disks and mini-magnets consisting of two single YBCO disks, respectively.\cite{9} In both cases, Ag was added to the YBCO disks which were placed into austenite Cr-Ni steel tubes. Previous attempts to combine the two beneficial effects of Ag and Zn additions failed due to the solubility of Zn in Ag in oxygen atmosphere. High trapped fields have also been reported for bulk Sm-123 with additions of Ag. At the surface of Sm-123 disks, trapped fields of 2.1\,T at 77\,K and of 8\,T at 40\,K have been observed.\cite{10}
 
In the present paper, the combined influence of Ag additions (improving the mechanical properties of YBCO) and of Zn doping (improving the pinning properties) on the maximum trapped field of single YBCO disks and of YBCO mini-magnets is studied. Furthermore, we report on attempts to obtain high trapped fields in the temperature range around 50\,K using Zn-doped YBCO disks with improved pinning behaviour. 

Melt textured YBCO bulk samples were prepared with varying content of Zn [YBCO(+Zn)] and with 0.12\,wt\,\% Zn combined with 10\,wt\,\% Ag [YBCO(+Ag/Zn)] using Sm-123 seed crystals. Details of the sample preparation are reported elsewhere.\cite{4,11,12} The applied melt process results in large-domain monoliths up to 50\,mm in diameter and about 12\,mm in height. The $a,b$-planes of the large grains were oriented parallel to the surface of the cylindrically shaped samples. The microstructure revealed small, homogeneously distributed Y$_{2}$BaCuO$_{5}$ precipitates 200-1000\,nm in size and, if Ag was added, supplementary inclusions of solidified Ag droplets in the range between 1 and 10\,$\mu$m in size. 

\begin{figure}[t]
\begin{center}
\includegraphics[height=5.5cm]{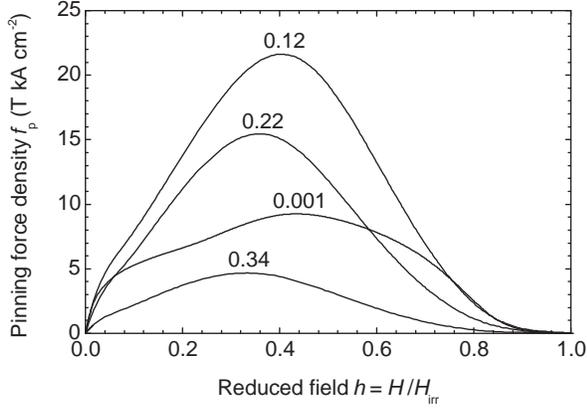}
\end{center}
\caption{Field dependence of the volume pinning force for Zn-doped YBCO samples in the range of zinc-concentrations between 0.001\,wt\,\% and 0.34\,wt\,\%.}
\label{figure1}
\end{figure}

At 77\,K, the disks were characterized by field mapping of the trapped field on the surface of the samples using an axial Hall sensor with an active area of 400\,$\mu$m$^{2}$. A maximum trapped field of $B_{\rm 0}$ = 1.2\,T was achieved in zinc-doped YBCO disks 35\,mm in diameter, whereas typical values of $B_{\rm 0}=0.8\,$T were found in undoped YBCO disks of the same size. The improved pinning properties of zinc-doped YBCO were confirmed by magnetization measurements on small samples (diameter 3\,mm). The volume density of the pinning force $f = j_{\rm c} B$ of several zinc-doped YBCO samples determined from magnetization-vs.-field curves at\,T = 75\,K is plotted in Fig.\,\ref{figure1} versus the reduced magnetic field $h =H/H_{\rm irr}$ with $H_{\rm irr}$ as the irreversibility field. The maximum pinning force density $f_{\rm max}$ is found to increase with increasing zinc concentration $x$ in the range of small concentrations $x< 0.12$\,wt\,\% and to decrease for higher concentrations. The improvement of the pinning force by zinc impurities occupying Cu sites in the CuO$_{2}$ planes can be attributed to pair breaking by locally induced magnetic moments in the neighborhood of the zinc impurities. The size of these disturbed regions within the CuO$_{2}$ planes was found\cite{13} to coincide approximately with the coherence length of YBCO and, therefore, they can be considered as effective pinning defects at low Zn concentrations. A strong $T_{\rm c}$ suppression is observed\cite{14} if these disturbed regions around the Zn impurities begin to overlap for higher Zn concentrations. The subsequent decrease of $f_{\rm max}$ for Zn concentrations $x>0.12$\,wt\,\% (see Fig.\,\ref{figure1}) is caused by this $T_{\rm c}$ suppression. Therefore a Zn concentration of 0.12\,wt\,\% was chosen to investigate the temperature dependence of the trapped fields for YBCO(+Zn). The same Zn concentration was also used for the investigated YBCO(+Ag/Zn) disks. 

\begin{figure}[b]
\begin{center}
\includegraphics[height=5.5cm]{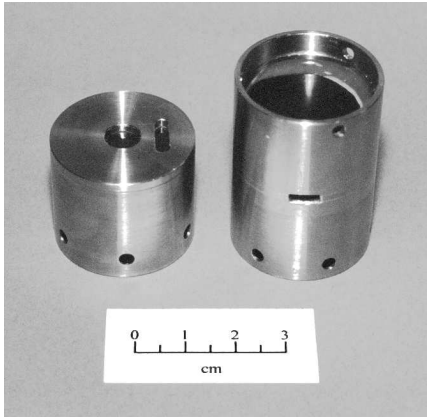}
\end{center}
\caption{Steel tubes with embedded YBCO permanent magnets. Left: single disk; right: mini-magnet. The notches are used to mount the Hall sensor and the thermometer.}
\label{figure2}
\end{figure}

\begin{figure}[t]
\begin{center}
\includegraphics[height=5.5cm]{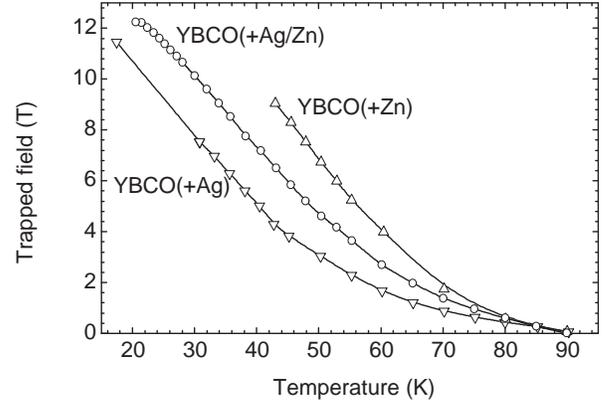}
\end{center}
\caption{Temperature dependence of the maximum trapped field $B_{\rm 0}$ measured on top of Ag-doped \protect\cite{9}, Zn-doped and Ag/Zn-doped YBCO single disks.}
\label{figure3}
\end{figure}

\begin{figure}[b]
\begin{center}
\includegraphics[height=5.5cm]{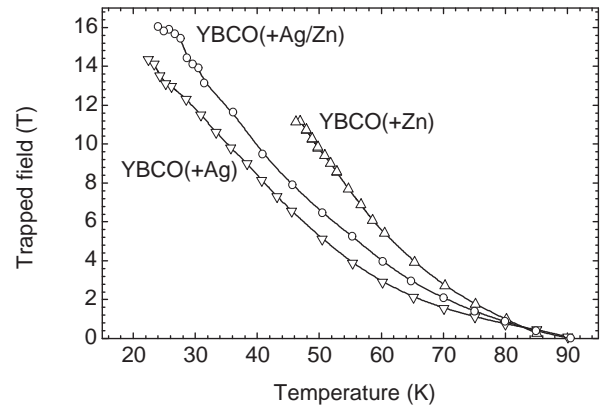}
\end{center}
\caption{Temperature dependence of the maximum trapped field $B_{\rm 0}$ in the gap of mini-magnets consisting of two Ag- \protect\cite{9}, Zn- and Ag/Zn-doped YBCO disks.}
\label{figure4}
\end{figure}

The temperature dependence of the maximum trapped field $B_{\rm 0}$ was measured on the top of single YBCO disks as well as in the gap between two disks. The measurements were performed in the variable temperature cryostat of a superconducting 18\,T magnet. The cylindrical pellets were cut to diameters of 24\,mm for YBCO(+Zn) and 22\,mm for YBCO(+Ag/Zn) and embedded in austenite Cr-Ni steel tubes with wall thicknesses of 3\,mm and 4\,mm, respectively (see Fig.\,\ref{figure2}). The small gap between the steel tube and the YBCO disks was filled with epoxy using a vacuum impregnation technique. For the investigation of single disks, an axial Hall sensor (AREPOC HHP-SA) was positioned on top of the disks in the centre, whereas a transversal Hall sensor (AREPOC HHP-MP) was placed in the gap between the two disks of the investigated mini-magnets. The size of the gap was 2.2\,mm (for the YBCO(+Zn) mini-magnet) and 2.6\,mm (for the YBCO(+Ag/Zn) mini-magnet). Additionally, a Cernox temperature sensor was fixed on top of one sample in each case. Both, magnetic field and sample temperature, were continuously recorded in steps of 1s during the measurements. After applying an external field at 100\,K, the temperature was reduced with a rate of 2.5\,K/min to the measuring temperature. When the thermal equilibrium was reached, the external field was ramped down at 0.1\,T/min to minimize heating of the samples by dissipative flux motion. Nevertheless, an increase in sample temperature of 0.2\,K at 60\,K and 1.8\,K at 25\,K was observed. The trapped fields $B_{\rm 0}$ presented in this paper (see Figs. \ref{figure3} and \ref{figure4}) correspond to the field value recorded immediately after reaching an external field of zero. Afterwards, the disks were heated up to 100\,K in order to remove the remanent field. To avoid thermally induced flux jumps the heating rate was limited to 0.5\,K/min below 24 K, to 1\,K/min between 24\,K and 30\,K and to 5\,K/min above 30 K.


The temperature dependence of the trapped field $B_{\rm 0}$ in YBCO single disks and mini-magnets with different combinations of Ag and Zn is shown in Figs. \ref{figure3} and \ref{figure4}. The $B_{\rm 0}$(T) curves for YBCO(+Zn) show a strong increase of the trapped field with decreasing temperature. Compared with the earlier investigated YBCO disks with 10\,wt\,\% Ag addition [YBCO(+Ag)]\cite{9}, the improved pinning properties of this material result in a considerable shift of the $B_{\rm 0}$(T) curves to higher temperatures by about 18\,K (single disks) and 15\,K (mini-magnets). Maximum $B_{\rm 0}$ values of 9\,T (at 44\,K) and 11.2\,T (at 47\,K) were obtained in YBCO(+Zn) single disks and mini-magnets, respectively, before cracking occured. The addition of Ag to YBCO was found to result in a reduced density of microcracks in the material\cite{15} and, therefore, increases the tensile strength and the achievable trapped field. The highest trapped fields were obtained for YBCO(+Ag/Zn) disks showing improved mechanical and pinning properties. In particular, trapped fields up to 16\,T (at 24 K) were achieved in YBCO(+Ag/Zn) mini-magnets. As expected, the corresponding $B_{\rm 0}$(T) curves (for single disks and mini-magnets) lie between those for YBCO(+Zn) and YBCO(+Ag). For the YBCO(+Ag/Zn) disks, cracking was observed at higher $B_{\rm 0}$ values than for the YBCO(+Ag) disks. This can be explained by the improved reinforcement of the YBCO(+Ag/Zn) disks for which the wall thickness of the steel tubes was enlarged to 4\,mm compared to 2\,mm in the case of the YBCO(+Ag) disks. \cite{9} 

Mostly, cracking of the YBCO disks was observed during magnetizing. However, especially in the temperature range between 20\,K and 30\,K the disks were damaged by flux jumps that occur during the heating procedure after the measurement for removing the trapped field. Despite the low rate used to warm up the samples, flux jumps due to thermomagnetic instabilities were observed for high trapped fields. The features of a small self-terminating flux jump are shown in the upper panel of Fig.\,\ref{figure5}: %

\begin{figure}[tb]
\begin{center}
\includegraphics[height=7cm]{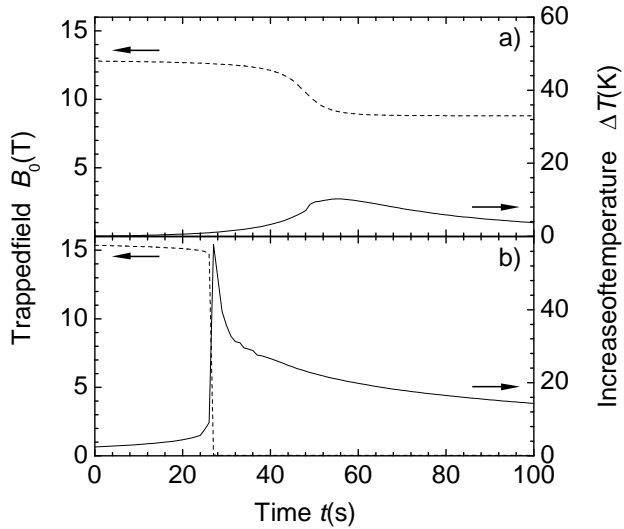}
\end{center}
\caption{Trapped field (- - -) and increase of temperature (---) versus time for YBCO minimagnets. a) YBCO(+Ag) with $B_{\rm 0}\approx 13\,$T; b) YBCO(+Ag/Zn) with $B_{\rm 0}\approx 15\,$T.}
\label{figure5}
\end{figure}%
This partial flux jump released only a small part of the magnetic energy stored in the YBCO disk. The corresponding increase of the temperature by about 10\,K caused no degradation of the YBCO disk. On the other hand, the complete flux jump shown in the lower panel of Fig.\,\ref{figure5} caused a rapid change of the entire magnetic energy into thermal energy turning the superconductor to the normal state. This flux jump leads to an irreversible damage of the YBCO disk.

In conclusion, very high magnetic fields up to 16\,T were trapped in YBCO(+Ag/Zn) disks embedded in a steel tube. These YBCO bulk magnets with improved mechanical and pinning properties resist the strong tensile stress acting during the magnetizing process on the superconductor, because the tensile stress is partly compensated by the compressive stress exerted by the steel tube. In order to remove the trapped field, the YBCO disks have to be warmed up. Thereby, below 30\,K, cracking of the disks may occur due to flux jumps.

The authors are indebted to B. Preu\ss \, and P. Bartusch for their help in sample characterization. This work was supported by the German Bundesministerium f\"{u}r Bildung und Forschung under contract \#13N7677.

\end{document}